# Electrothermally Modulated Nanophotonic Waveguide-integrated Ring Resonator

*Sujal Gupta*[α,*] *and Jolly Xavier*[α,β,*]

[α]*OPC, Indian Institute of Technology Delhi, New Delhi-110016, Delhi, India;*

[β]*SeNSE, Indian Institute of Technology Delhi, New Delhi-110016, Delhi, India;*

---

**ABSTRACT:** Reconfigurable integrated chips of photonic components and networks are envisaged to play a key role in realizing highly efficient integrated photonic information processing. Electrothermally modulated optical effect (ETMOE) is a powerful thermo-optic tuning mechanism for silicon photonic devices, enabling precise optical control via localized Joule heating. We present a rigorous and three-dimensional electronic-photonic co-integrated approach with the nonlinear numerical coupling of temperature and wavelength-dependent material properties to comprehensively model ETMOE in silicon waveguides and resonators. A platinum-based symmetric heater is optimized using advanced true 3D numerical simulations, achieving efficient ETMOE-based tuning while mitigating asymmetric heat distribution. In addition to a complete design and analysis of the fully integrated three-dimensional switch, we also evaluate single-mode waveguide cutoff, heater-to-waveguide separation, heater dimensions, and thermal dissipation, providing a framework for ETMOE-based optimization. The findings contribute to energy-efficient, programmable photonic systems for neuromorphic computing, optical interconnects, and reconfigurable photonic networks.

---

## INTRODUCTION

Silicon photonics has facilitated the development of compact, high-performance photonic integrated circuits (PICs) for telecommunications[1–3], optical computing[4–6], and sensing[7,8]. Thermo-optic (TO) tuning has emerged as a key method for reconfiguring photonic devices due to its compatibility with complementary metal-oxide-semiconductor (CMOS) fabrication processes and broad wavelength tunability[5,6,9,10]. However, the limited electronic-photonic co-simulation platforms and reliance on simplistic heater designs often overlook the nonlinear temperature and wavelength dependence of material properties, limiting accuracy and efficiency in their analysis[11]. Electrothermally modulated optical effect (ETMOE) leverages localized Joule heating to modulate the refractive index of optical waveguides through the TO effect, enabling precise wavelength control. Microheaters integrated with silicon waveguides have been extensively investigated for applications such as optical filters[1,3], modulators[10,12,13], and switches[14,15]. Studies have demonstrated effective thermal tuning using metallic[9,10,12,13,16] and doped silicon[6,15] microheaters due to their high stability and compatibility with silicon-on-insulator (SOI) platforms. Despite these advancements, existing approaches primarily rely on simplified thermal models, neglecting critical nonlinearities in material properties, such as temperature-dependent electrical conductivity, heat capacity, and temperature and wavelength-dependent refractive index, leading to spatial variations in optical behavior.

Micro-ring resonators (MRRs) are widely used in optical switching[2,14,15,17], wavelength filtering[1,3], and neuromorphic computing[4] due to their high-quality factors and compact footprint. However, the thermal sensitivity of MRRs presents a significant challenge, as undesired temperature variations can lead to resonance drifts and optical crosstalk in densely integrated photonic circuits[5,6,11]. Prior research has explored passive thermal desensitization techniques, such as self-heating compensation and thermally isolated waveguides, to mitigate these issues [18]. Nevertheless, active thermal tuning remains necessary for precise wavelength locking in reconfigurable photonic networks[6]. Recent studies have proposed machine-learning-based control mechanisms for thermal tuning[19]. In contrast, others have investigated novel heater architectures, such as optimized metallic microheaters for high-speed operation[9,10,13,16] and cascaded double-ring architectures for enhanced tuning range in sensing applications[8]. However, these approaches do not fully incorporate the nonlinearities in thermal-optical interactions, potentially limiting their accuracy in real-world conditions.

We present a complete rigorous electronic-photonic co-integrated approach with nonlinear coupling of material properties for a comprehensive numerical analysis of ETMOE using advanced 3D simulations. A complete methodology is developed, encompassing the physics of device-level optimization—including waveguide, heater, and ring resonator design—to application-level implementation of phase modulator and switch. The proposed approach is well-suited for programmable nanophotonic systems, enabling precise control and calibration for emerging photonic applications.

## DESIGN APPROACH AND METHODOLOGY

The schematic in Fig. 1(i) displays the double bus (DB)-racetrack resonator (RTR) with a symmetric heater positioned sideways above the racetrack ring waveguide. The optical paths (red) represent the signal flow through the waveguide, while the electrical pathways (black) connect to the heater input and output electrical pads. This configuration enables electro-thermal tuning of the racetrack's refractive index for effective optical modulation. The cross-sectional view in Fig. 1(ii) provides detailed information about the material layers that can be used in the fabrication process. The modulator can be fabricated on a SOI wafer, where the optical waveguides consist of a silicon (Si) core surrounded by silicon dioxide ($SiO_2$). The heater, made of platinum (Pt), is positioned atop the waveguide. To ensure proper adhesion of Pt to $SiO_2$, a thin titanium (Ti) adhesive layer is used. The electrical pads are made of gold (Au), which adheres effectively to Pt, but for high-frequency applications, tungsten (W) vias are introduced to ensure improved performance and connectivity. The dashed box highlights the key section of the device which is used to achieve accurate simulation results. Additionally, the inset presents a 3D schematic of a simplified straight waveguide for ETMOE, emphasizing the functional arrangement of the layers. The meshed structure, derived from the schematic in Fig. 1(i), is presented for finite element method (FEM)-based advanced numerical simulation and analysis in Fig. 1(iii). Meshing ensures accurate resolution of electrical, thermal, and optical, interactions across the multilayered device. The meshing parameters are optimized for the device's operation at a wavelength (λ) of 1550 nm. The minimum mesh size is calculated as λ/(15×n), while the maximum mesh size is λ/(7×n), where n is the refractive index of the respective material at this wavelength. Finer mesh elements are applied to areas of high complexity, such as the optical waveguide and the heater region, to enhance computational accuracy while a coarser mesh is used in the reason not of interest.

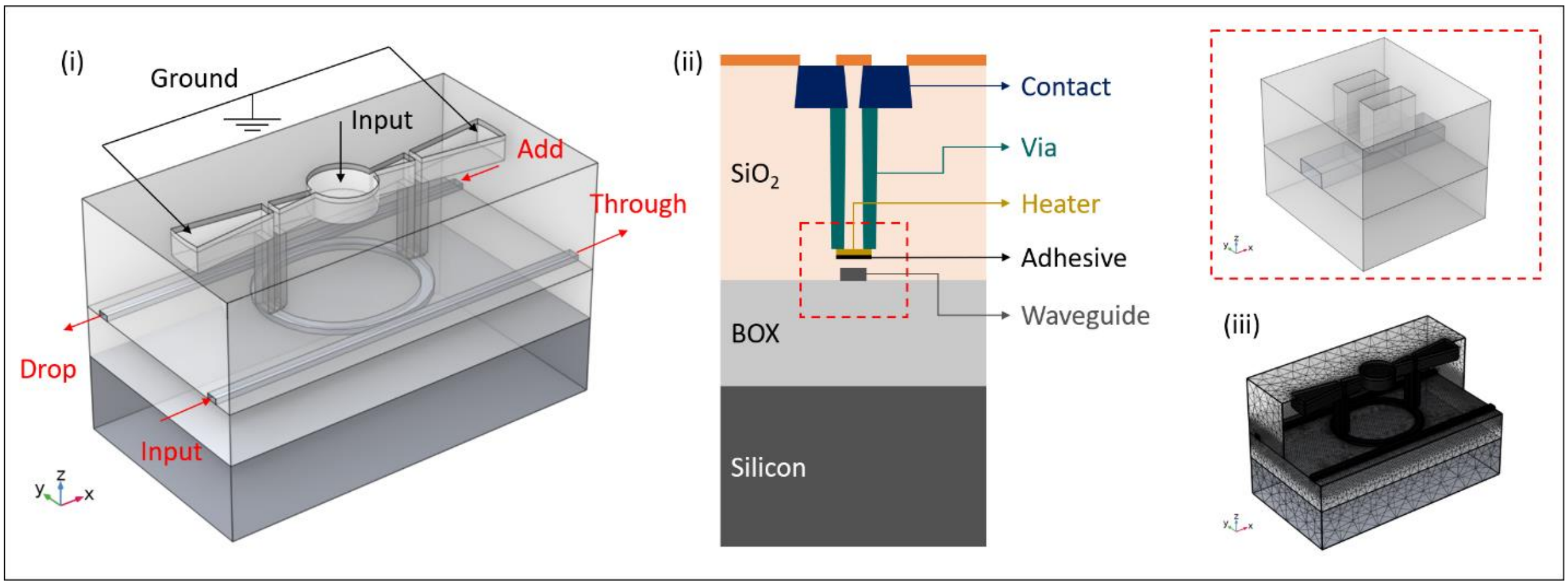

**Figure 1. Schematic and meshing of the ETMOE based devices.** (i) Schematic representation of the ETMOE in DB-RTR, with optical signals (red) propagating through the waveguide and electrical connections (black) linked to a symmetric heater placed atop the racetrack. (ii) Cross-sectional view of the device layers used for simulation and can be utilized for fabrication. The dashed red box highlights the key section of the device that is sufficient for accurate simulation. The inset shows a schematic of the ETMOE in straight waveguide. (iii) FEM-based meshing of the DB-RTR structure depicted in (i).

Fig. 2 outlines the theoretical and computational foundation for the ETMOE achieved through the coupling of the system's electromagnetic (EM) heating, heat transfer, and TO behavior[6,10,17,20]. This comprehensive setup is implemented in COMSOL Multiphysics using electric current (EC), heat transfer in solids (HTS), and EM waves frequency domain (EWFD) physics interfaces. The Joule heating mechanism forms the basis of the temperature rise in the heater. The EC interface solves for the current density:

$$\mathrm{J} = \sigma \mathrm{E} + \mathrm{J_e}, \mathrm{E} = -\nabla V$$

where J is the total current density, σ is the temperature-dependent electrical conductivity, E is the electric field, $J_e$ is the externally generated current density, and V is the electric scalar potential.

This heating generates a volumetric heat source $Q_{j,v} = Q_e = \mathrm{J}.\mathrm{E}$. This term is passed via the EM heating Multiphysics coupling as the direct EM heat source $Q_e = Q$ to the HTS interface which solves the heat diffusion equation:

$$\rho C_p \mathrm{u}. \nabla T + \nabla . q = Q_e + Q_{ted}$$

incorporating temperature-dependent material properties such as density (ρ), specific heat ($C_p$), and thermal conductivity (k). Here, T is the temperature, and q is the conductive heat flux given by $q = -k\nabla T$. The term $Q_{ted}$ represents thermoelastic damping (mechanical energy losses in the material) and is absent ($Q_{ted}$=0). The temperature distribution obtained from the HTS interfac modifies the wavelength- and temperature-dependent refractive index and extinction coefficient (κ) of the materials, as described by the TO coefficients $\frac{\partial n_0}{\partial T}$ and $\frac{\partial \kappa_0}{\partial T}$ :

$$n = n_0 + \frac{\partial n_0}{\partial T} \Delta T, \kappa = \kappa_0 + \frac{\partial \kappa_0}{\partial T} \Delta T$$

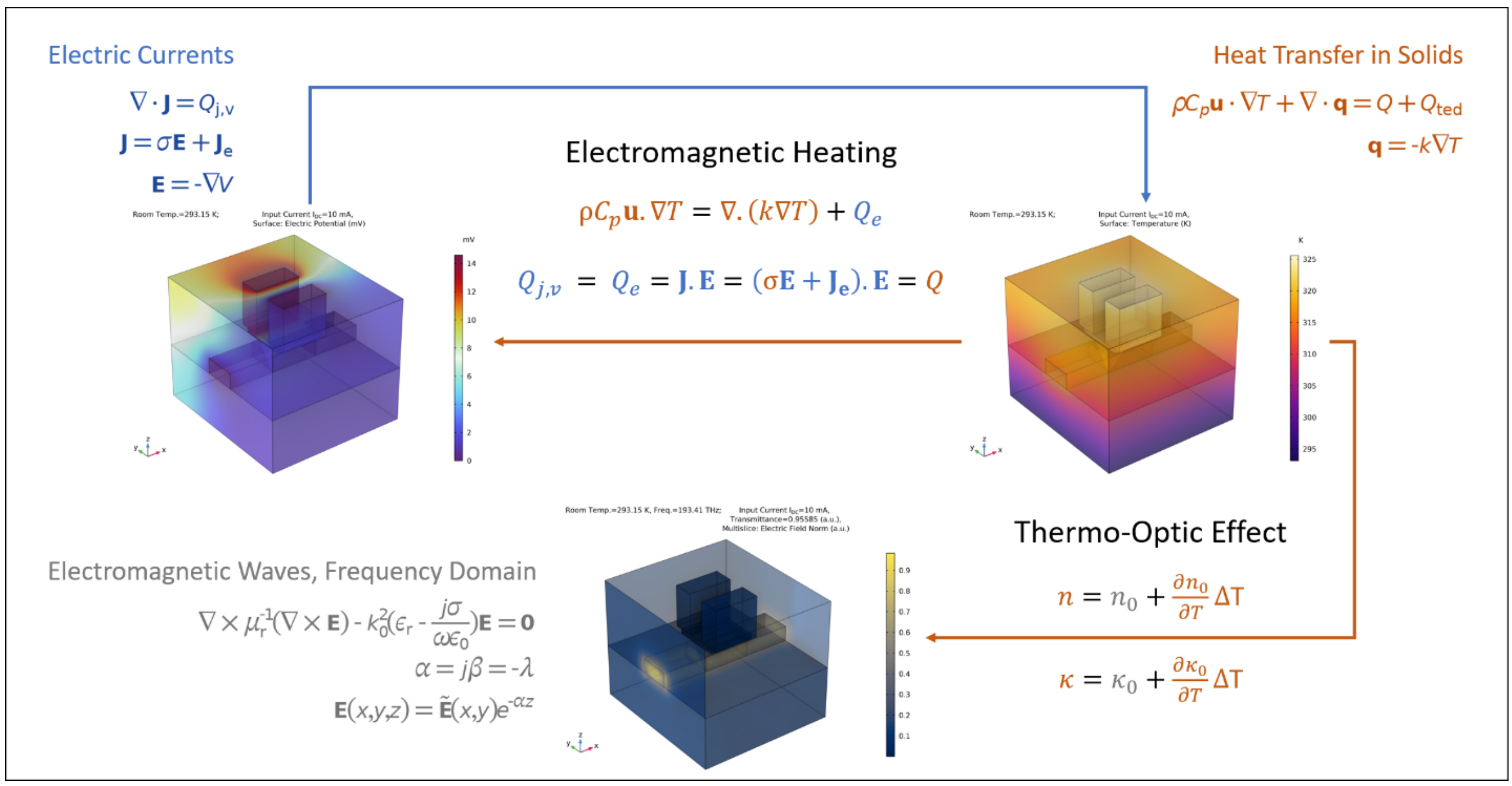

**Figure 2. Theoretical framework for ETMOE.** Key physical equations governing EM heating (current density, heat transfer, and thermal flux) and the TO effect (mode propagation and refractive index tuning) are presented, highlighting the temperature and wavelength-dependent properties used in advanced numerical analysis.

This change, accounted by the self-developed TO Multiphysics coupling between HTS and EWFD interfaces, impacts the optical propagation constant (β) and attenuation constant ($\alpha = j\beta$). The EWFD interface solve for light propagation in the system, incorporating the spatially varying refractive indices caused by TO-effect. The governing wave equation solved in this interface is expressed as:

$$\nabla \times \frac{(\nabla \times \mathrm{E})}{\mu_r} - k_0^2 \left(\epsilon_r - \frac{j\sigma}{\omega\varepsilon_0}\right) \mathrm{E} = 0$$

where $\mu_r$ is the material's relative permeability, $k_0$ is the wave number in a vacuum, and $\varepsilon_r$ is the material's relative permittivity, ω is the angular frequency, and $\varepsilon_0$ is the permittivity in a vacuum. This setup ensures that all relevant physics are accurately captured in a single electronic-photonic co-simulation environment, enabling a comprehensive analysis of ETMOE.

## RIGOROUS ANALYSIS AND RESULTS

### I. Mode Analysis and Heater Position

Fig. 3(i) evaluates the guiding properties of the Si waveguide by analyzing the effective mode index and power confinement for different waveguide widths, keeping the height fixed at 220 nm as it supports pure transverse electric (TE) mode at a wavelength of 1550 nm in a slab waveguide. The study shows the first three quasi-mode excitations: fundamental TE ($TE_0$), fundamental transverse magnetic ($TM_0$), and first TE ($TE_1$). The waveguide supports single $TE_0$ mode below the cutoff width of 480 nm. A width of 460 nm is selected to ensure single-mode operation, effectively eliminating higher-order modes like $TE_1$. This choice ensures predictable and efficient performance, making the waveguide suitable for most photonic applications. The $TE_0$ mode profile in Fig. 3(ii) exhibits strong confinement, with an effective mode index of 2.3738 and a power confinement of 77.1%. In contrast, the $TM_0$ mode profile in shows weaker confinement, with an effective mode index of 1.7411 - 0.0011i and a power confinement of only 41.2%. These results confirm the waveguide's compatibility with $TE_0$-dominant applications, highlighting the trade-offs in mode confinement for different polarizations with an attenuation constant of 1.03 dB/cm.

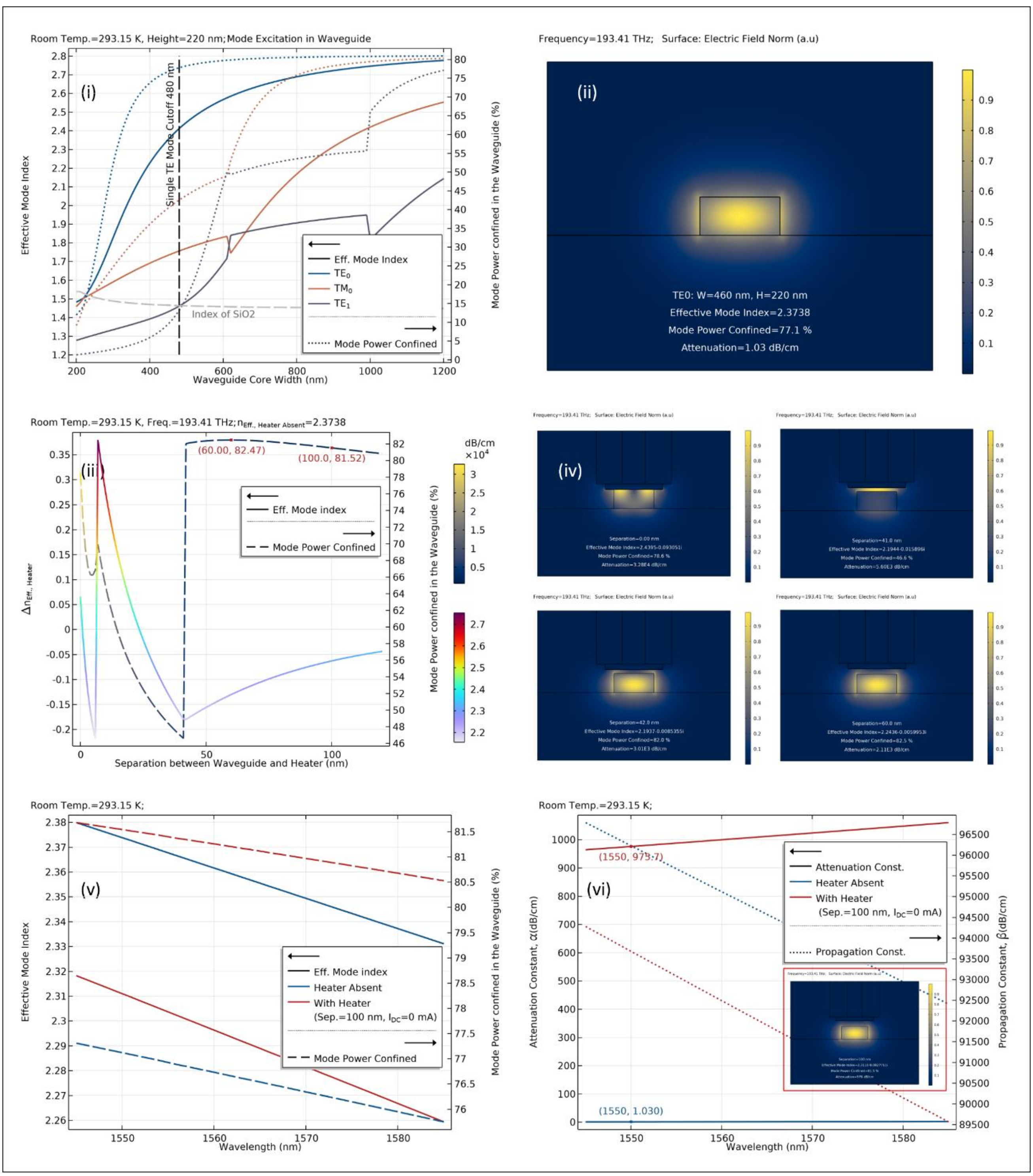

**Figure 3: Comprehensive mode analysis for the Si waveguide.** *Absence of heater:* (i) Mode excitation analysis showing the effective mode index (solid lines, left y-axis) and power confinement percentage (dotted lines, right y-axis) as a function of waveguide width for a fixed waveguide height of 220 nm. The first three quasi-modes, $TE_0$ (blue), TM0 (brown), and $TE_1$(grey), are analyzed showing single TE mode cutoff of 480 nm. (ii) $TE_0$ mode field profiles for the chosen waveguide dimensions (width = 460 nm, height = 220 nm), with effective mode index of 2.3738, power confinement of 77.1% and attenuation constant of 1.03 dB/cm. *Presence of heater:* (iii) Mode confinement analysis illustrating the change in effective mode index (solid line, left y-axis) and percentage power confinement (dashed line, right y-axis) as functions of the separation between the waveguide and the top heater with no input current. Color maps on the right side illustrate the change in attenuation constant (top) and group index (bottom). (iv) $TE_0$ mode field profiles for various heater separations (0 nm, 41 nm, 42 nm, and 60 nm), showing the restoration of proper confinement at separations ≥42 nm. (v, vi) Modal dispersion analysis comparing the effective mode index (solid lines, left y-axis) and power confinement (dashed lines, right y-axis) in (v) and attenuation constant (solid lines, left y-axis) and propagation constant (dotted lines, right y-axis) in (vi)

as functions of wavelength in the absence of a heater (blue) and with the heater placed 100 nm above the waveguide (brown). The inset in (vi) shows the $TE_0$ mode field profiles for heater separation of 100 nm (red box).

Fig. 3(iii) explores the impact of the heater's proximity to the waveguide (referred as separation) on mode confinement. The results reveal that when the heater is directly on the waveguide (0 nm separation), the mode profile is distorted despite ~79% power confinement within the waveguide (Fig. 3(iv), top-left). As the separation increases to 6 nm and 7 nm, the shifts in the effective mode index shows oscillatory response (negative to positive, Fig. 3(iii)). Below a 42 nm separation, confinement drops (Fig. 3(iii) and Fig. 3(iv), top-right). However, stable behavior is achieved at a 42 nm separation (Fig. 3(iv), bottom-left) and beyond, with power confinement exceeding 80% in the waveguide (Fig. 3(iii)). The best confinement is observed at a 60 nm separation (Fig. 3(iv), bottom-right). However, a 100 nm separation is chosen for practical fabrication robustness, balancing fabrication tolerances, and operational reliability. This trade-off results in only ~1% less power confinement than the optimal case at a separation of 60 nm. Fig. 3(v-vi) illustrates the spectral performance of the waveguide without and with the heater positioned 100 nm above it. As the wavelength increases, the effective mode index and power confinement exhibit characteristic dispersion (Fig. 3(v)) with corresponding attenuation constant and propagation constant (Fig. 3(vi)). The comparison underscores the heater's perturbative effects, emphasizing the need for precise design to minimize losses or distortions. The inset in Fig. 3(vi) shows the $TE_0$ mode profile exhibits strong confinement, with an effective mode index of 2.3110-0.0027i and a power confinement of 81.5% with attenuation constant of 976 bB/cm at wavelength of 1550 nm.

## II. ETMOE-based Heater Optimization for Straight Waveguide Phase Modulator

Fig. 4 analyzes the ETMOE of a straight silicon waveguide modulator integrated with a straight Pt nanoheater of thickness 60 nm and width 660 nm at a wavelength of 1550 nm. The ETMOE exploits localized Joule heating to induce refractive index modulation in the waveguide (the portion beneath the heater referred as shadowed waveguide, considered to present further analysis) via the TO effect, enabling efficient optical signal modulation. Fig. 4(i) examines the effect of heater-to-waveguide separation on ETMOE while maintaining a fixed heater length of 660 nm. As the separation increases, transmittance improves due to reduced heat dissipation within the waveguide, lowering thermal gradients across the core and minimizing scattering and absorption losses. This is evident in the attenuation constant, which reduces from 2007.51 dB/cm at 60 nm (optimum case) to 980.45 dB/cm at 100 nm, at an input current of 10 mA. However, excessive separation weakens thermal coupling, limiting modulation efficiency. The color maps on the right illustrate variations in the attenuation constant (top) and electric potential distribution (bottom). A 100 nm separation was chosen to balance effective modulation, fabrication tolerance, and optical performance. Fig. 4(ii) explores ETMOE behavior for varying heater lengths, keeping the separation at 100 nm. Transmittance initially increases with heater length due to improved thermal interaction with the waveguide. However, beyond a range of 600–800 nm, transmittance stabilizes, indicating diminishing returns in modulation efficiency. The selected heater length of 660 nm ensures sufficient heating to effectively modulate a single optical wavelength.

Fig. 4(iii) utilizes the selected heater dimensions (660 nm length and width, 60 nm thickness, 100 nm separation) to evaluate ETMOE as a function of input current. Transmittance increases with rising current, corresponding to the heater's temperature increase, which remains below the melting points of the device materials. These results confirm the suitability of the proposed design for practical ETMOE applications. Fig. 4(iv) presents the ETMOE field distribution at 10 mA input current, depicting the electric potential, temperature, and mode profiles. These results highlight the localized heating effect of the nanoheater and its efficient coupling with the waveguide to achieve phase modulation at operational wavelength of 1550 nm. Finally, Fig. 4(v) evaluates the time-dependent steady-state heating response for input voltages of 5 mV, 10 mV, and 50 mV. The temperature rise and terminal current profiles reveal a steady-state response time of 7 μs. However, the thermal response of the shadowed waveguide varies based on the applied voltage, indicating different heat transfer dynamics. Fig. 4(vi) extends this analysis to pulsed input voltages with a fixed pulse width of 20 μs and amplitudes of 10 mV for the temperature rise in the shadowed waveguide. The steady-state rise times of 7 μs and fall times of 12 μs, are observed. While the rapid rise and fall time of ~4 μs is observed, considering the 10% and 90% lines. These results indicate that while pulsed operation is feasible, modulation speed is ultimately constrained by the fall time, which is governed by the thermal properties of the materials. A similar analysis is applicable for the ETMOE and thermal analysis of a sectored heater designed for thermal modulation in a racetrack resonator.

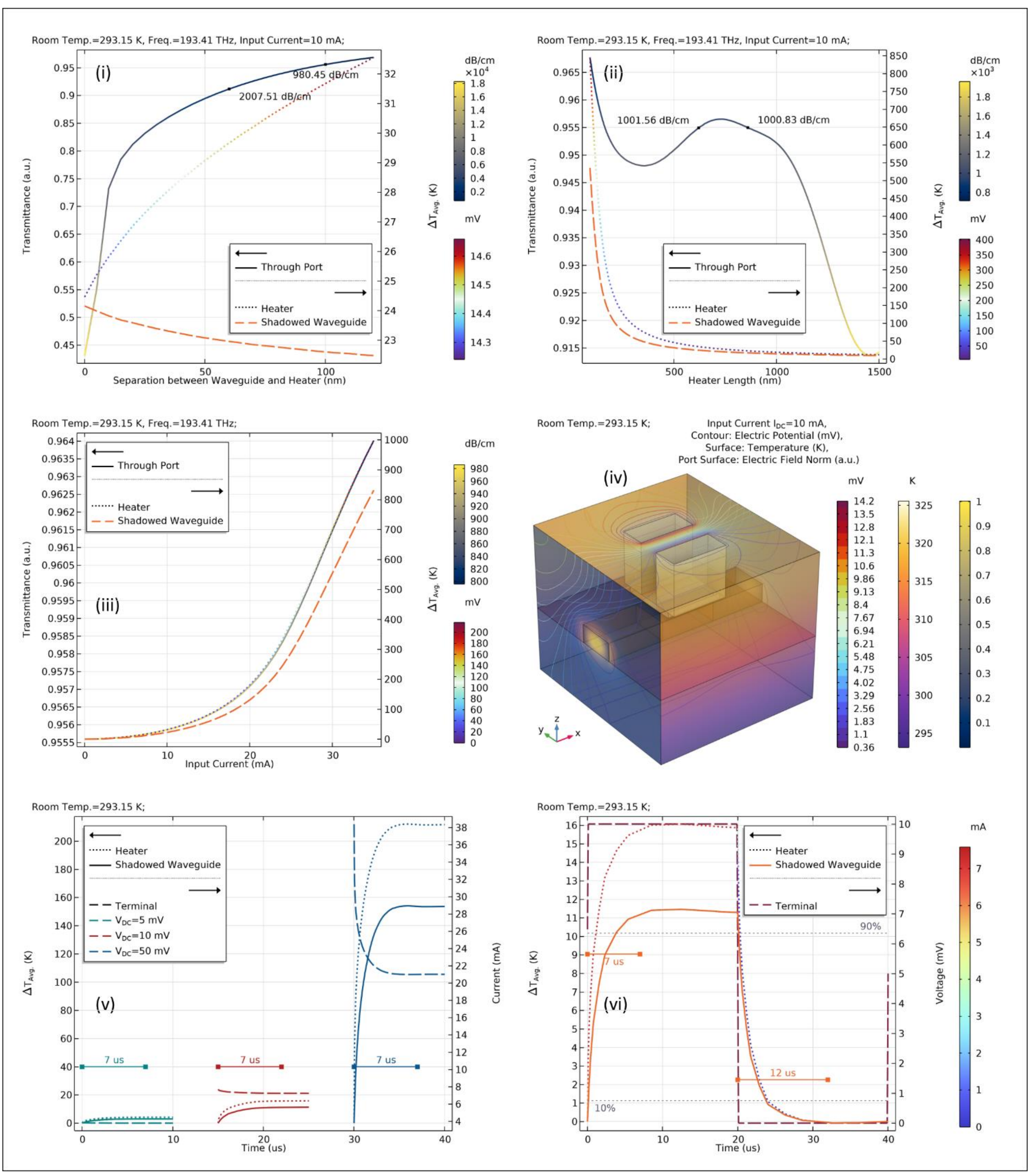

**Figure 4. ETMOE analysis.** (i–ii) Transmittance response (solid line, left y-axis) of a straight waveguide modulator as a function of temperature (dotted and dashed lines, right y-axis) for varying structural parameters with a fixed heater width of 660 nm and thickness of 60 nm at a fixed input current of 10 mA and an optical wavelength of 1550 nm: (i) heater-to-waveguide separation with a fixed heater length of 660 nm, (ii) heater length with a fixed separation of 100 nm. (iii) ETMOE based response of the waveguide for varying input current, showing transmittance (solid line, left y-axis) and temperature changes (dotted and dashed lines, right y-axis) with selected heater dimensions (660 nm length, 660 nm width,

and 60 nm thickness), placed at a separation of 100 nm. Color maps on the right side of (i–iii) illustrate the variation in attenuation constant (top) and electric potential magnitude (bottom). (iv) ETMOE based field distribution at 10 mA input current, showing electric potential, temperature, and $TE_0$ mode profile. (v) Time-dependent heating response for constant input voltages of 5 mV (teal), 10 mV (maroon), and 50 mV (blue), illustrating the temperature rise (solid and dotted lines, left y-axis) and terminal current (dashed lines, right y-axis) with a steady-state response time of 7 µs. (vi) Time-dependent heating response for pulsed input voltages with a fixed pulse width of 20 µs and pulse amplitudes of 10 mV, illustrating the temperature change (solid and dotted lines, left y-axis) and terminal voltage (dashed lines, right y-axis), with steady-state rise times of 7 µs and steady-state fall times of 12 µs for the temperature change in the shadowed waveguide.

## III. Racetrack Resonator, Resonance Tuning and, Tuneable Photonic Switch

The wavelength-dependent transmission spectrum of the RTR in Fig. 5(i) showcases sharp resonance dips at 1550 nm and 1581.51 nm, separated by a free spectral range (FSR) of 31.5 nm. The resonator achieves an impressive quality (Q)-factor of 2583 with full-width half-maxima (FWHM) of 0.6 nm, reflecting its high optical performance. The dimensions of the RTR are numerically optimized for critical coupling, including a coupling length (l) of 660 nm, a racetrack radius (R) of 2.6949 µm, and a gap of 145 nm. While optimization numerical precision of up to 100 pm for the radius and 5 nm for the gap ensures a resonance dip of approximately -30 dB. These parameters satisfy the theoretical RTR resonator resonance conditions described by the equation:

$$m\lambda = 2(\pi R + l)n_{eff}$$

where m is the resonance order determining the number of wavelengths that fit around the resonator and $n_{eff}$ is the effective refractive index. The $n_{eff}$ accounts for waveguide dispersion and field confinement within the silicon waveguide, enabling precise control of resonance conditions. The high Q-factors of 2583 highlight the resonator's energy storage capability and minimal optical losses, attributed to optimized waveguide geometry and low scattering and absorption in the Si and SiO2 materials. These characteristics ensure prolonged photon lifetime and substantial field enhancement, which are critical for optical filtering and dense wavelength-division multiplexing applications. The FSR, observed as the spacing between adjacent resonance peaks, depends on the group index ($n_g$) and the optical path length ($L_{Optical}$) of the resonator, given by:

$$FSR = \frac{\lambda^2}{n_g L_{Optical}}$$

This relationship implies that reducing the racetrack radius increases the FSR, enabling closely spaced devices with reduced crosstalk. Fig. 5(ii) showcase the $TE_0$ field distribution at 1550 nm, demonstrating efficient light confinement and resonance within the racetrack geometry.

Integrating a symmetric sectored platinum heater onto the arms of the RTR ring enables precise thermal control of the device (Fig. 5(iii)). This placement ensures uniform temperature distribution, mitigating asymmetric refractive index gradients that could degrade performance. The modified dimensions for critical coupling—a radius of 2.6968 µm and a gap of 80 nm—maintain a resonance dip of approximately -30 dB. Finally, resonance shifts of 0.15 nm, 0.55 nm, and 1.1 nm, observed under applied voltages of 10, 20, and 30 mV, respectively, relative to the 0 mV reference, confirm robust and predictable TO tuning. Fig. 5(iv) show the ETMOE field distribution for an input current of 10 mA. The device's flexibility allows calibration to specific wavelengths without re-optimizing after integrating the sectored heater. This combination of precise tuning, compact size, and high performance makes it good for PIC in programmable and dense optical interconnects.

Fig. 5(v, vi) highlights the performance of a tunable ETMOE based DB-RTR switch designed for compact and precise wavelength-selective switching. The device exploits the TO effect, wherein localized heating induced by an applied voltage modulates the refractive index within the resonator. Fig. 5(v) demonstrates the wavelength-dependent switching behavior of the device under different control voltages. At a reference wavelength of 1549.74 nm for through and 1549.66 nm for drop port, two distinct states are observed. In the OFF/ON state at 0 mV, the transmission at the through port is 0.1415 (OFF), while the drop port shows a transmission of 0.3881 (ON). In the ON/OFF state at 50 mV, the through port transmission increases to 0.8274

(ON), while the drop port decreases to 0.0791 (OFF). This switching behavior illustrates the device's versatility, as it can function as either a transmission-based switch (through-port dominant) or a drop-based switch (drop-port dominant), depending on the specific application requirements.

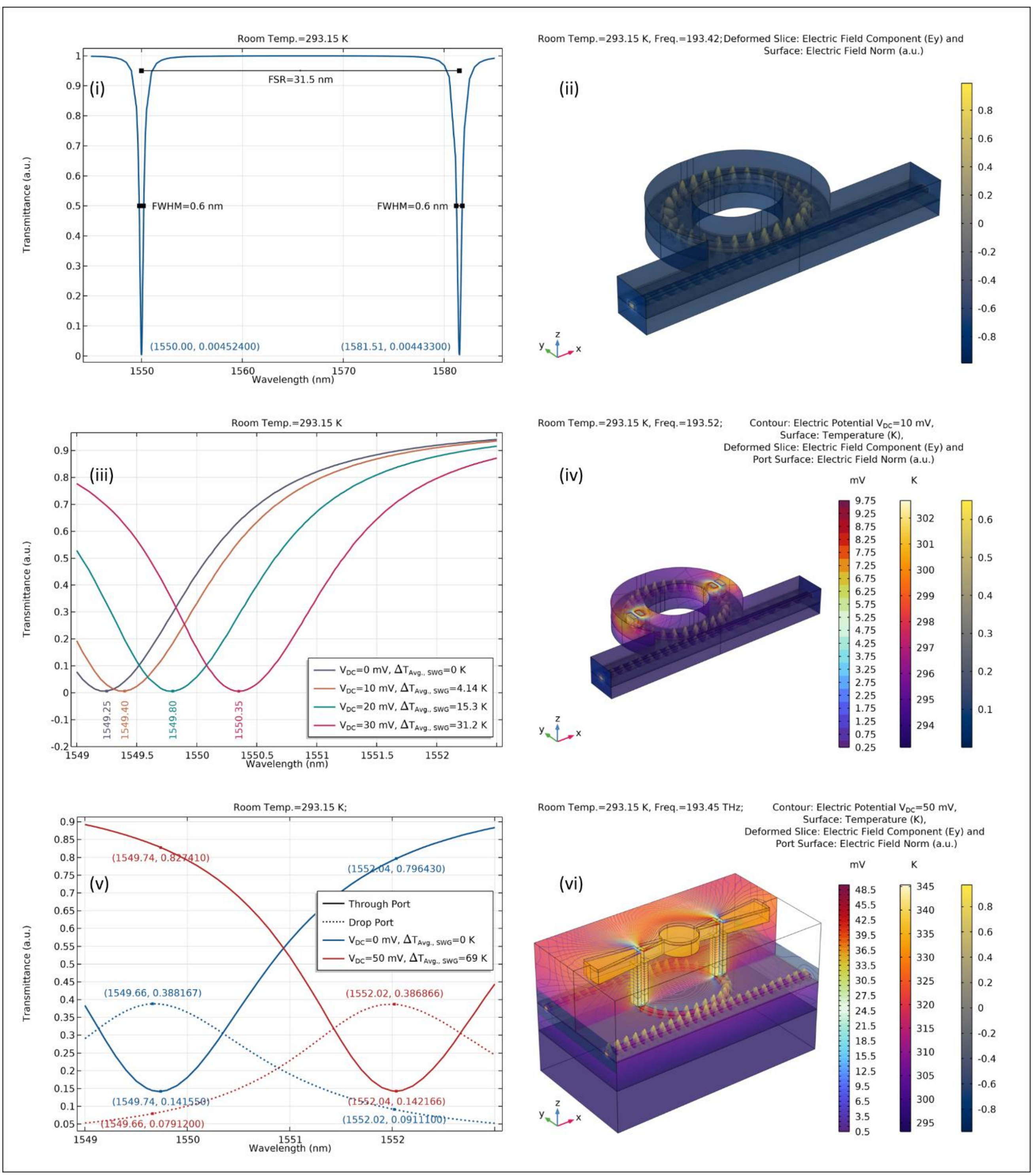

**Figure 5. EM resonance analysis and ETMOE based applications.** *Single-bus-RTR:* (i) Wavelength-dependent transmission spectrum of the RTR. (ii) The $TE_0$ field distribution at resonance wavelength of 1550 nm. (iii) ETMOE based tuning of the wavelength-dependent transmission spectrum for input voltages of 10, 20, and 30 mV. The corresponding TO resonance shifts of 0.15, 0.55, and 1.1 nm are observed relative to the 0 mV reference. (iv) The ETMOE tuned field distribution for an input current of 10 mA. *ETMOE based DB-RTR switch:* (v) Wavelength-dependent transmission spectra for control input voltages of 0 mV (off/on state) and 50 mV (on/off state), showing distinct switching behavior at specific wavelengths of 1549.74 nm. Switching occurs as a function of the control voltage, with transitions observed at through-port/drop-port transmissions. (vi) The field distribution of the ETMOE based DB-RTR switch corresponds to an input voltage of 50 mV.

A similar trend is observed at a wavelength of 1552.04 nm (through port) and 1552.02 nm (drop port), showcasing the switch's robust and predictable performance over a defined wavelength. Compared to simpler microring designs, the compact racetrack geometry provides enhanced thermal confinement and localised refractive index gradients, resulting in precise spectral responses. Fig. 5(vi) presents the field distribution corresponding to an input voltage of 50 mV. The field is tightly confined within the RTR, demonstrating efficient energy coupling between the waveguides and the resonator. This field distribution underscores the effectiveness of the design in practical scenarios, ensuring low insertion losses and strong coupling efficiencies. The consistent field profile also validates the resonator's suitability for large-scale photonic integration. Unlike many designs with high power consumption or limited wavelength tunability, the demonstrated switch achieves precise modulation over multiple wavelengths while maintaining a low-voltage operation.

## DISCUSSION AND OUTLOOK

The performance of ETMOE-based photonic devices critically depends on the interplay between electrical, thermal, and optical properties. Studies on thermo-optic tuning have primarily focused on microheater design and material selection[10,12,13,15]. However, many of these studies rely on simplified linear models that fail to account for temperature-dependent variations in material properties in Joule heating and temperature and wavelength-dependent properties in TO-effects, leading to discrepancies between simulation and experimental results. In contrast, our work integrates an electronic-photonic co-simulated, fully nonlinear coupling approach in multilayered devices, ensuring greater accuracy in predicting device behavior under real-world conditions.

A key advantage of our study is the comprehensive parametric optimization of heater-to-waveguide separation and heater dimensions. We showed that a 100 nm separation balances modulation efficiency and fabrication feasibility. Such optimizations are rarely explored in prior works, where fixed heater dimensions are often assumed without extensive analysis of their impact on optical performance.

Our study shows a steady-state response time of 7 μs, the previous control and calibration work reports 12 μs[9]. For pulsed inputs, we achieve a consistent steady state rise time of 7 μs and a fall time of 12 μs for the temperature increase in the shadowed waveguide using a 20 μs pulse width and 10 mV pulse amplitude. Previous studies reported fall times exceeding 4 μs[13], our optimized heater and waveguide design also achieves a rapid rise and fall time of less than 4 μs when 10% and 90% lines are considered for the temperature change in the shadowed waveguide. However, while the rise time is comparable to state-of-the-art microheaters, the fall time remains a bottleneck, dictated by the inherent thermal relaxation properties of $SiO_2$ cladding and the buried oxide layer. Future efforts should focus on engineering thermally conductive pathways or integrating advanced thermal dissipation structures.

Researchers have demonstrated high-Q optical filtering and switching capabilities using thermal tuning for microring and racetrack resonators[2,3,17]. However, thermal crosstalk and resonance stability remain significant concerns in densely integrated photonic circuits[16,18]. Our study presents a racetrack resonator design with a Pt-based symmetric sectored heater configuration, enabling precise electrothermal control while mitigating asymmetric thermal gradients. The observed resonance shifts of 0.15 nm, 0.55 nm, and 1.1 nm for applied voltages of 10, 20, and 30 mV showcase the robustness of our nonlinear simulation model. The applied voltage can go up to 200 mV for improved resonance shifts, although extra care should be taken to avoid resonance crossing/overlapping in integrated devices.

Traditional TO tuning methods often require high power consumption, making them less suitable for energy-efficient photonic networks. Some recent studies have explored machine-learning-based feedback control for heater optimization[19], while others have investigated alternative materials, such as nonvolatile phase-change materials (PCMs), to reduce power dissipation[15]. However, PCMs introduce optical losses and reliability concerns due to repeated phase transitions. In contrast, our approach maintains compatibility with conventional silicon photonics while achieving precise and repeatable tuning with lower power requirements. We have also rigorously investigated a fully electronic-photonic co-simulated realistic ETMOE-based 3D switch, showing feasibility for practical silicon photonic applications. Unlike prior designs focused on theoretical configurations or simplified 2D models, our 3D switch accounts for fabrication constraints and real-world operational conditions. The optimized design minimizes insertion loss, enhances switching contrast, and ensures thermal stability, making it suitable for high-speed photonic interconnects and signal routing applications.

ETMOE-based tuning is particularly relevant for neuromorphic photonics[4] and reconfigurable optical interconnects[5], where precise wavelength control is essential. While some studies have implemented integrated control loops for thermal stabilization, our work lays the groundwork for fully automated, calibration-free tuning by providing an accurate numerical framework for predictive modeling. Our study advances ETMOE-based tuning by incorporating nonlinear material coupling, optimizing heater design, and achieving competitive modulation speeds. While the thermal relaxation time remains a limiting factor, our approach provides a scalable and energy-efficient solution for next-generation silicon photonics applications.

## CONCLUSION

Our work advances the electronic-photonic co-integrated approach for ETMOE-based tuning in silicon photonics by introducing a fully nonlinear numerical model that captures temperature and wavelength-dependent variations in material properties. Through comprehensive parametric optimization, we showed that a 42 nm heater-to-waveguide separation or more is sufficient for single-mode confinement, while a 100 nm separation balances modulation efficiency and fabrication feasibility. The state-of-the-art competitive modulation speeds are shown with a stable rise time of 7 µs and a fall time of 12 µs for the temperature change in the shadowed waveguide under pulsed operation. Additionally, our study highlights the fundamental thermal limitations in silicon photonics, particularly the extended fall time dictated by material relaxation properties. Future research should explore thermally conductive pathways or alternative material integration to mitigate these constraints. The optimized symmetric heater integrated with RTR enables precise tuning with minimal power consumption, achieving resonance shifts of 1.1 nm for a control voltage of 30 mV, which can be operated at higher voltage below the melting point of materials for increased resonance shifts. Furthermore, we have designed and analyzed a fully integrated realistic 3D switch, showing its feasibility for practical silicon photonic applications, ensuring optimized insertion loss, switching contrast, and thermal stability. The insights from this study lay the foundation for next-generation programmable photonic systems, offering scalable solutions for dynamic wavelength control in optical networks and neuromorphic photonics for more efficient photonic devices by bridging the gap between theoretical modeling and practical implementation.

### ASSOCIATED CONTENT

**Supporting Information (available on request)**. Material Properties and Multiphysics Setup provides detailed insights into the temperature- and wavelength-dependent material properties used in the advanced numerical simulations analysis. Additionally, it outlines the Multiphysics setup of the ETMOE, developed by coupling electromagnetic heating with the thermo-optic effect explained in the theory. (Material-Properties_Multiphysics-Setup_Results-Supporting-Information). Several gif animations files were generated for supporting thorough understanding of obtained results. (Animation 1-5)

### AUTHOR INFORMATION

**Corresponding Author**
Sujal Gupta- OPC, Indian Institute of Technology Delhi, New Delhi-110016, Delhi, India;
Email: opz228317@opc.iitd.ac.in
Jolly Xavier- SeNSE, Indian Institute of Technology Delhi, New Delhi-110016, Delhi, India;
Email: jxavier@sense.iitd.ac.in

**ACKNOWLEDGMENT**

S.G. gratefully acknowledges the award of the INSPIRE Fellowship [DST/INSPIRE/03/2021/001134] from the Department of Science and Technology, India.

**BRIFES.** *Advanced electronic-photonic co-integrated 3D numerical analysis of electrothermally modulated optical effects from physics to device level application.*